\documentclass[doublespacing]{elsart}

\usepackage{natbib}

\usepackage{graphics}
 \usepackage{graphicx}
 \usepackage{epsfig}

\usepackage{amssymb}

\begin{document}

\begin{frontmatter}



\title{The structure and properties of vacancies in Si nano-crystals calculated 
by real space pseudopotential methods}


\author{S. P. Beckman\corauthref{cor1}\thanksref{label1}\thanksref{label2}}
\ead{spbeckman@gmail.com}\ead[url]{http://sbeckman.net/scott/}
\author{James R. Chelikowsky\thanksref{label1}}
\corauth[cor1]{corresponding author}

\address[label1]{Center of Computational Materials, Institute of 
Computational Engineering and Sciences, Departments of Physics and 
Chemical Engineering, University of Texas, Austin, TX 78712 USA.}
\address[label2]{Department of Physics and Astronomy, Rutgers 
University, Piscataway, NJ 08855 USA.}

\begin{abstract}
The structure and properties of vacancies in a 2 nm Si nano-crystal are studied 
using a real space density functional theory/pseudopotential method.
It is observed 
that a vacancy's electronic properties and energy of formation are directly 
related to the local symmetry of the vacancy site.  The formation energy for 
vacancies and Frenkel pair are calculated.  It is found that both defects have 
lower energy in smaller crystals.  In a 2 nm nano-crystal the energy to form 
a Frenkel pair is 1.7 eV and the 
energy to form a vacancy is no larger than 2.3 eV.  The energy barrier for 
vacancy diffusion is examined via a nudged elastic band algorithm.  
\end{abstract}


\end{frontmatter}

\pagebreak

\section{Background}
As the size of micro-electronic devices continues to diminish 
the devices are approaching the nano-scale~\cite{chau2003}.  
Although scaling improvement is still possible within the 
traditional micro-electronics 
paradigm,~\cite{chau2003,walls2003} 
there is an effort to move to devices constructed directly from 
nano-structures~\cite{lieber2007,liang2007}.  Nano-devices 
can be constructed that function both to fill the roll of 
existing technology~\cite{liang2007,yang2007} and as novel devices 
whose performance is dependent upon nano-induced 
properties~\cite{lieber2007,stall2005}.  
\par
One aspect of device design and manufacture that requires 
special attention is doping.  In micro-electronics the 
placement and control of dopants has been studied in depth 
and is continuing to be studied~\cite{queisser1998}.  At the 
nano-scale dopants behave very differently than in bulk.  
Quantum confinement limits the possible 
concentration of some impurities due to an increase in the enthalpy 
of formation.  This is the 
so-called \emph{self-purification} effect in 
nano-structures~\cite{GD1}.  However, by carefully controlling 
the growth environment it is possible to create selectively doped 
nano-structures~\cite{liang2007,yang2007,erwin2005,lu2006}.  It 
is unclear whether the dopants that are incorporated during the 
growth process remain in the active region of the nano-structure 
because they are kinetically limited, or if they are 
thermodynamically stabilized by a yet unidentified effect.  
\par
The diffusion of impurities depends strongly 
upon their interaction with intrinsic defects, 
especially interstitial atoms and vacancies.  At the nano-scale it is 
anticipated that quantum confinement will affect the energy of 
formation and migration of intrinsic defects.  In addition, the free 
surfaces, which can move to reduce internal stresses, will also influence 
the properties of defects.  Understanding diffusion at the 
nano-scale requires understanding the balance between these two effects.  
\par
The density functional theory/pseudopotential methods used here are 
encoded in the software \textsc{parsec}~\cite{kohn1965,JRC1}.  The real space 
grid is chosen to be 0.5 a.u.\footnote{The atomic units of length 
used here are Bohr radii unless specified otherwise.}, which is 
sufficient for force and energy 
convergence.  Aperiodic boundaries conditions are employed with a 
vacuum region of greater than 6 \AA~between the surface of the 
nano-crystal and the cell boundary.  
The local 
density approximation is used for the exchange and correlation 
functional~\cite{ceperley1980}.
Troullier-Martins 
pseudopotentials~\cite{troullier1993} are used with Si having 
a valence of $3s^{2}3p^{2}3d^{0}$ and a cutoff radius 
of 2.5 a.u.\ for all angular channels.  The hydrogen 
pseudopotential has a valence of $1s^{1}$ and a 2.0 a.u.\ cutoff.  
The pseudopotential is transformed into local and non-local 
components by the Kleinman-Bylander~\cite{kleinman1982} 
transformation with the p channel selected as the local component 
for Si.  All atomic structures are optimized until the forces 
are smaller than 0.04 eV/\AA.  
\par
To calculate the diffusion barriers a nudged elastic band 
method is implemented~\cite{henkerman2000}. 
The velocity Verlet algorithm is used to update the images in 
configuration space.  The vacuum region is increased by 
2 \AA~to insure that the nano-crystal does not interact with the 
boundaries.   
The spring coefficient of the elastic band is initially set to a 
large value, to insure the stability of the calculation, but as the 
elastic band approaches the minimum energy path, the coefficient is 
reduced to improve the convergence rate.  
\par
\begin{figure}[e!]
\centering
\includegraphics [width=2.5in] {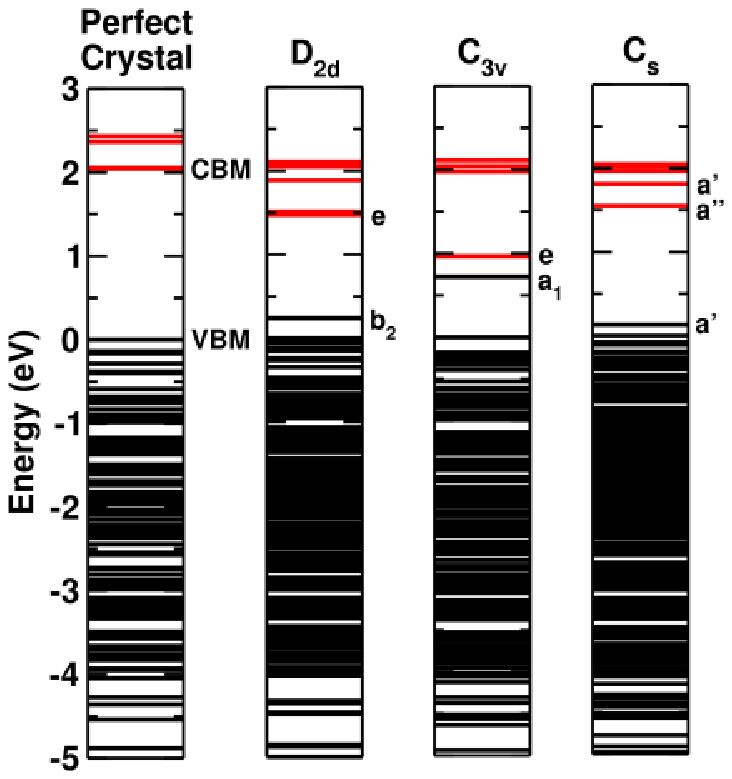}
\caption{The electronic states of vacancies with various local symmetries.  
The occupied states are shown in black and the empty states in red. 
The representation that each defect state belongs is written 
to the right of the corresponding state.\label{2nmstatemap}}
\end{figure}

A 2 nm diameter, hydrogenated, Si nano-crystal $(Si_{175}H_{116})$ is 
examined.  The geometric center of the nano-crystal corresponds to an 
Si atom.  Each surface Si is bonded to no more than 2 H.  A hydrogen passivated 
surface represents the situation where the surface is the most capable 
of relaxation.  It is anticipated that replacing the hydrogen with oxygen, polymers, or an Si 
reconstruction will lead to a slight constriction of the surface.  The 
calculated gap in a defect free Si nano-crystal is 2.04 eV, as 
shows in Fig.~\ref{2nmstatemap}.  
\par
\section{Results and Discussion}
\begin{figure}[e!]
\centering
\includegraphics [width=2.5in] {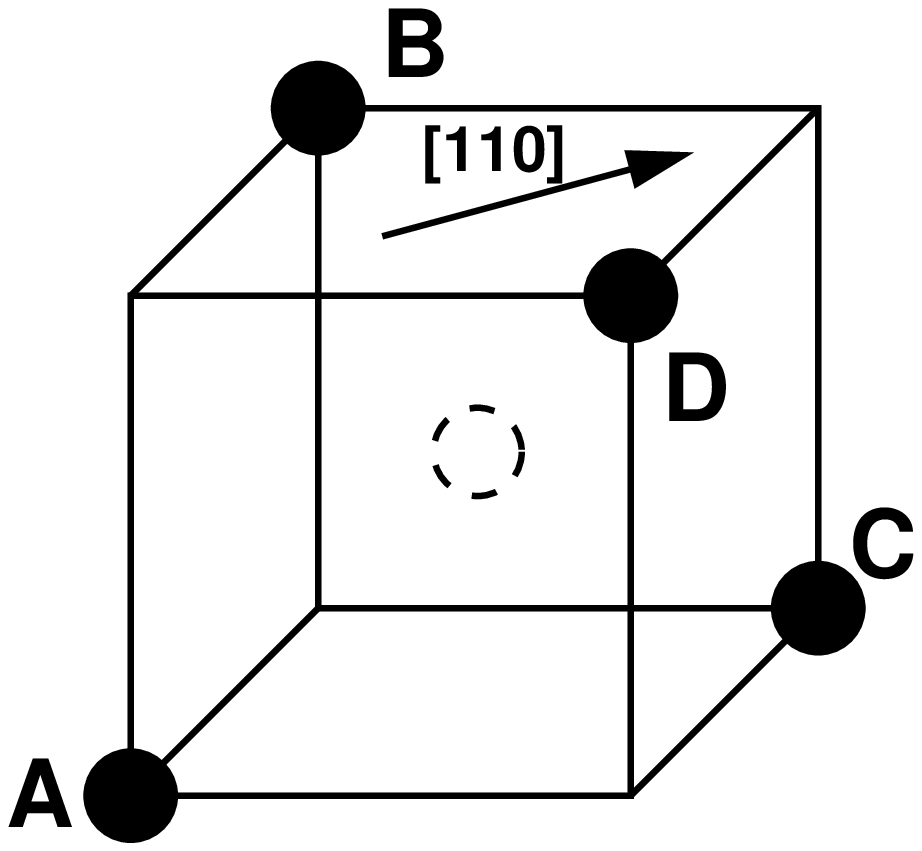}
\caption{A geometric representation of the local structure around a vacancy.\label{tet}}
\end{figure}

A vacancy is placed at sites along the [110] direction from the center of the 
crystal to the surface.  
The structures, states, and total energies 
are calculated for each configuration.  
When a vacancy is placed in the center of 
the nano-crystal, the local symmetry of the vacancy is that of the 
nano-crystal's point group, $T_{d}$.  The vacancy undergoes a 
Jahn-Teller distortion similar to that observed in bulk, which results 
in the $T_{d}$ symmetry being broken to $D_{2d}$~\cite{watkins1986,serdar2001}.
The reduction in local symmetry causes a reduction of degeneracies of 
the electronic states.  The $T_{d}$ configuration has a triply degenerate, 
partially occupied, state associated with the $t_{2}$ representation.  
This splits into a fully occupied state belonging to the $b_{2}$ representation 
and an empty, doubly degenerate, state belonging to the $e$ representation 
of the $D_{2d}$ point group.  The states that are localized at the defect 
are identified in Fig.\ \ref{2nmstatemap} as well as the representations to which 
they belong.  The energies of each of the structures are presented in 
Table \ref{etable}.  The geometry of the calculated structures are given in 
Table \ref{postable}, using the atomic structure shown in Fig.\ \ref{tet} as 
a reference.  
\par

\begin{table}[e!]
\begin{center}
\begin{tabular}{|c|c||c|c||c|c|}
\hline
\multicolumn{2}{|c|}{$D_{2d}$} & \multicolumn{2}{|c|}{$C_{3v}$} & \multicolumn{2}{|c|}{$C_{s}$}\\
\hline
\hline
\multicolumn{6}{|c|}{Defect States (eV)} \\
\hline
Rep. & Ene. & Rep. & Ene. & Rep. & Ene. \\
\hline
$b_{2}$ & 0.24 & $a_{1}$ & 0.72 & a'  & 0.13 \\
\hline
$e$     & 1.47 & $e$     & 0.96 & a'' & 1.55 \\
\hline
\multicolumn{4}{|l||}{}         & a'  & 1.81  \\
\cline{5-6}
\hline
\hline
\multicolumn{6}{|c|}{Total Energy (meV)} \\
\hline
\multicolumn{2}{|c|}{0} & \multicolumn{2}{|c|}{853} & \multicolumn{2}{|c|}{-329}\\
\hline
\hline
\end{tabular}
\caption{The energy of defect states and total energy of nano-crystals containing 
vacancies.  The defect levels are expressed in eV.  The highest occupied, non-localized, 
state is taken to 
be the valence band maximum (VBM) and is set to zero.  The energy gap of a vacancy free 
crystal is 2.04 eV.  The total energy of the vacancy structures are 
presented in meV 
and the $D_{2d}$ structure is arbitrarily set to zero.}
\label{etable}
\end{center}
\end{table}

\begin{table}[e!]
\begin{center}
\begin{tabular}{|c|c|c|c|}
\cline{2-4}
\multicolumn{1}{l|}{} & \multicolumn{3}{|c|}{Distances (\AA)} \\
\hline
Segment & $D_{2d}$ & $C_{3v}$ & $C_{s}$ \\
\hline
\hline
$\overline{AB}$ & 3.4 & 3.4 & 3.6 \\
\hline
$\overline{AC}$ & 2.7 & 3.4 & 2.5 \\
\hline
$\overline{AD}$ & 3.4 & 3.4 & 3.6 \\
\hline
$\overline{BC}$ & 3.4 & 2.9 & 3.4 \\
\hline
$\overline{BD}$ & 2.7 & 2.9 & 2.7 \\
\hline
$\overline{CD}$ & 3.4 & 2.9 & 3.4 \\
\hline
\end{tabular}
\caption{The local structure about vacancies with different symmetries.  The reference geometry 
is shown in Fig.\ \ref{tet} and the details of the orientation is given in the text.  Using this 
computational method the bond-length in the center of a vacancy free nano-crystal is 2.3 \AA, and the distance 
between the segments, is 3.8 \AA.}
\label{postable}
\end{center}
\end{table}

When the vacancy is located one site from the center, a Jahn-Teller distortion 
is still present, however, the symmetry of the distortion is different.  The 
local symmetry of the site depends not only upon the nano-crystal's 
point group, but also the location of the vacancy in the nano-crystal.  
Because the vacancy 
is positioned along one of the $T_{d}$ three-fold rotation axis, the local symmetry obeys 
this three-fold rotation, but most of the other operations of the $T_{d}$ 
point-group are destroyed.  The local 
symmetry 
about the vacancy is $C_{3v}$.  The geometry of the vacancy is given by
 Table \ref{postable} and Fig.\ \ref{tet}, where the center of the nano-crystal 
is site A.  The defect states are shown in Fig.\ \ref{2nmstatemap} where the 
occupied state belongs to the $a_{1}$ representation and the degenerate, empty, 
states belongs to the 
$e$ representation.  
\par
If the vacancy is located yet one site farther from the center, in the [110] 
direction, all of the 
$T_{d}$ symmetry operations are absent except for one mirror plane that 
contains both the center of the crystal and the vacancy site and has a normal in 
the $\left[\bar{1}10\right]$ direction.  The vacancy has symmetry $C_{s}$.  The 
geometry is given in 
 Table \ref{postable} and Fig.\ \ref{tet}, where the vector $\vec{AC}$ points from the 
center of the crystal toward the surface, in the [110] direction.  Because $C_{s}$ 
has no doubly degenerate, irreducible, representations, the defect states, shown 
in  Fig.\ \ref{2nmstatemap}, have no degeneracy.  
\par
The energies of the vacancy structures reported 
in Table \ref{etable} are directly related to the energies of the occupied 
defect states and the local symmetry. 
This is in agreement with published observations of the importance 
of symmetry in nano-crystals~\cite{GD2}.
If the diffusion 
mechanism for vacancies is limited to site hopping, then in these highly symmetric 
nano-crystals it is possible to trap vacancies at the center of the nano-crystal.  
It is also apparent that sites nearer the surface are lower in energy than 
sites in the interior.  When a vacancy is placed within two atomic layers of the 
surface, around 4 nm, the vacancy is spontaneously pulled to the surface.  
These calculations are in qualitative agreement with 
calculations performed on 3 nm Si nano-crystal~\cite{beckmanunpub}.  It is 
predicted that the effect 
of symmetry on the properties of point defects will also be observed in other 
nano-structures.  In particular it is anticipated that vacancies and other 
point defects in [111], [001], and [112] nano-wires should differ from one 
wire to the next.  
\par
The energy of formation for a Frenkel-pair is investigated in 1 and 2 nm 
nano-crystals.  Interstitial atoms are placed at tetrahedral sites adjacent 
to vacancies that have the $D_{2d}$ structure.  The energy of formation is 
calculated to be 0.63 and 1.70 eV for the 1 and 2 nm nano-crystals respectively.  
The calculated energy of formation for a Frenkel defect, in bulk, is 
around 7 eV~\cite{centoni2005}.  The bulk calculation uses a vacancy-interstitial 
separation of 8 \AA, but this is not possible 
in these nano-structures.  It is concluded from this calculation that as a 
nano-structure decreases in size, the energy to spontaneously create intrinsic 
point defects diminishes.  The energies calculated here, are small 
enough that one expects to find nano-crystals containing Frenkel defects in any 
reasonable sized ensemble of nano-structures.  In principle these Frenkel defects 
can act to generate vacancy-interstitial pair that may be pulled toward the surfaces, 
creating an internal stirring within the nano-crystal.  Additional studies 
must be performed to understand the impact of this result on diffusion.  
\par
Subtracting the calculated total energy of a nano-crystal with a vacancy from the 
total energy of a perfect nano-crystal yields the energy of formation 
for a vacancy plus the energy to add an Si atom to the computational 
cell.  The energy to add an atom is dependent upon 
the computational method.  It is more tractable instead to calculate 
the relative energy of formation by defining one 
vacancy structure as the energy zero.  By setting the energy of formation for a 
vacancy in the center of a 1 nm crystal as zero it is determined that the 
relative energy of formation 
for a vacancy in a 2 nm crystal is 1.4 eV.  For a 3 nm 
crystal the relative energy of formation is calculated to be no larger 
than 2.0 eV.  The energy of formation 
in bulk is around 3.7 eV~\cite{centoni2005}.  It is deduced that the 
absolute energy of 
formation for a vacancy can be no larger than 1.7 and 2.3 eV for 1 and 2 nm 
nano-crystals.  These calculations demonstrate the influence of 
free surfaces on the energy of formation for point defects in nano-structures. 
Although it is known that for some classes of defects the 
quantum confinement, self-purification effect is dominant~\cite{GD1}, 
this is not universally true.  
\par
To study the migration barrier for vacancies, a 2 nm crystal centered on a Si-Si 
bond is examined.  The nudged elastic band method is employed to determine the 
minimum energy path to move a vacancy across the center of the crystal.  
The end configurations have $C_{3v}$ symmetry.  
The middle of the transition 
corresponds to placing the atom that is exchanging sites with the vacancy directly in 
the center of the crystal, creating a six-fold coordinated structure with $D_{3d}$ 
symmetry and bond lengths of 2.63 \AA.  
The energy of this structure is -0.64 eV relative to the $C_{3v}$ 
structure.  
The end configurations are only local minima. 
The lowest energy configuration is found between the $C_{3v}$ configuration and 
the $D_{3d}$.  The minimum has low symmetry and a relative 
energy of -1.24 eV.  There is no degeneracy in the defect states and the 
energy difference between the empty and full state is 1.1 eV.  
The vacancy in this configuration is much nearer to the center of the crystal, and 
is off of the [111] three-fold rotation axis.  
Although this arrangement introduces strain to the surrounding bonds, the surfaces 
help to reduce the strain energy.  This result suggests that the vacancy diffusion 
process may occur through pathways that do not require the vacancy to sit directly
upon atomic sites.  The barrier to moving across the center of the crystal from 
one minimum to another is 0.6 eV.  This is slightly smaller than the literature 
value for bulk Si, 0.8 eV~\cite{shimizu2007}.  
This indicates that the diffusivity of vacancies in Si nano-structures is enhanced.  

This work was supported in part by the National Science Foundation under
DMR-0551195 and the US Department of Energy under
DE-FG02-06ER15760 and DE-FG02-06ER46286.
Computational resources were provided in part by the National Energy Research 
Scientific Computing Center (NERSC) and the Texas Advanced Computing 
Center (TACC). 

\bibliographystyle{prsty}
\bibliography{XPO_40bib}

\end{document}